\documentclass[11pt]{extarticle}
\usepackage[english]{babel}
\usepackage{graphicx}
\usepackage{framed}
\usepackage[top=1 in,bottom=1in, left=0.8 in, right=0.8 in]{geometry}

\usepackage[utf8]{inputenc} 
\usepackage[T1]{fontenc}    
\usepackage{hyperref}       
\usepackage{url}            
\usepackage{booktabs}       
\usepackage{amsfonts}       
\usepackage{amsmath}
\usepackage{amsthm}
\usepackage{amssymb}
\usepackage{mathtools}
\usepackage{commath}
\usepackage{mathrsfs}
\usepackage{bbm}
\usepackage{nicefrac}       
\usepackage{microtype}      
\usepackage{lipsum}
\usepackage{graphicx}
\graphicspath{ {./images/} }
\usepackage{xcolor}
\usepackage[titletoc,toc,page]{appendix}

\usepackage{enumitem}  
\usepackage{comment}
\usepackage{float}
\usepackage[font=small,skip=2pt]{caption}

\theoremstyle{plain}

\newtheorem*{theorem*}{Theorem}

\newtheorem{lemma}{Lemma}

\theoremstyle{definition}

\newcommand{\x}{\mathbf{x}}
\newcommand{\M}{\mathbf{M}}

\newcommand{\R}{\mathbb{R}}

\newcommand{\E}{\mathbb{E}}

\newcommand{\Vm}{V_{\mathrm{m}}}

\newcommand{\X}{\mathbf{X}}

\newcommand{\rd}{\mathrm{d}}
\newcommand{\cD}{\mathcal{D}}

\newcommand{\bxi}{\boldsymbol{\xi}}
\newcommand{\bzeta}{\boldsymbol{\zeta}}

\title{Spikes can transmit neurons' subthreshold membrane potentials}
\date{}
\begin{document}
\author{Valentin Schmutz\footnote{UCL Queen Square Institute of Neurology, University College London, WC1E 6BT London, UK. Email:~v.schmutz@ucl.ac.uk}}
\maketitle
\begin{abstract}
    Neurons primarily communicate through the emission of action potentials, or spikes. To generate a spike, a neuron’s membrane potential must cross a defined threshold. Does this spiking mechanism inherently prevent neurons from transmitting their subthreshold membrane potential fluctuations to other neurons? We prove that, in theory, it does not. The subthreshold membrane potential fluctuations of a presynaptic population of spiking neurons can be perfectly transmitted to a downstream population of neurons. Mathematically, this surprising result is an example of concentration phenomenon in high dimensions.
\end{abstract}

Neurons in the mammalian brain mainly communicate through the emission of large, pulse-like depolarizations of their membrane potential called spikes. To fire a spike, a neuron's membrane potential ($\Vm$) needs to cross a threshold. Between the emission of spikes, the subthreshold membrane potential fluctuations of a neuron are, by definition, not transmitted to other neurons, but they carry precious information as they reflect the total synaptic input received by the neuron. 

The richness of sensory and behavioral information contained in the subthreshold membrane potential dynamics, compared to that contained in the timings of spikes, has been revealed by whole-cell $\Vm$ recordings of individual neurons in behaving animals \cite{Pet17}. For example, during whisking in mice, 
the subthreshold membrane potentials of pyramidal neurons in the layer 2/3 of the whisker primary somatosensory cortex (wS1) closely track whisker position which oscillates at a frequency of about $10$ Hz \cite{CroPet06}, 
while the same pyramidal neurons fire sparse spikes, most of them having a firing rate lower than $1$ Hz and some of them remaining completely silent \cite{CroPet06,PouPet08}. 
This type of evidence has put forward the idea that spikes are just ``the `tip of the iceberg' in terms of neuronal activity'' \cite{Pet17}. But if a single presynaptic neuron does not transmit any information about its subthreshold membrane potential fluctuations, does this imply that subthreshold information is lost, in the sense that it is not accessible to a postsynaptic neuron? At the single-neuron level, the loss is real; optimal estimation of potential from spikes only allows for partial recovery of the membrane potential \cite{PfiDay09,PfiDay10}. Adopting a network-level perspective, we prove that the membrane potential fluctuations of a presynaptic population of neurons emitting sparse spikes can be fully and perfectly transmitted to a postsynaptic population of neurons.

\begin{figure*}
    \includegraphics[width=\textwidth]{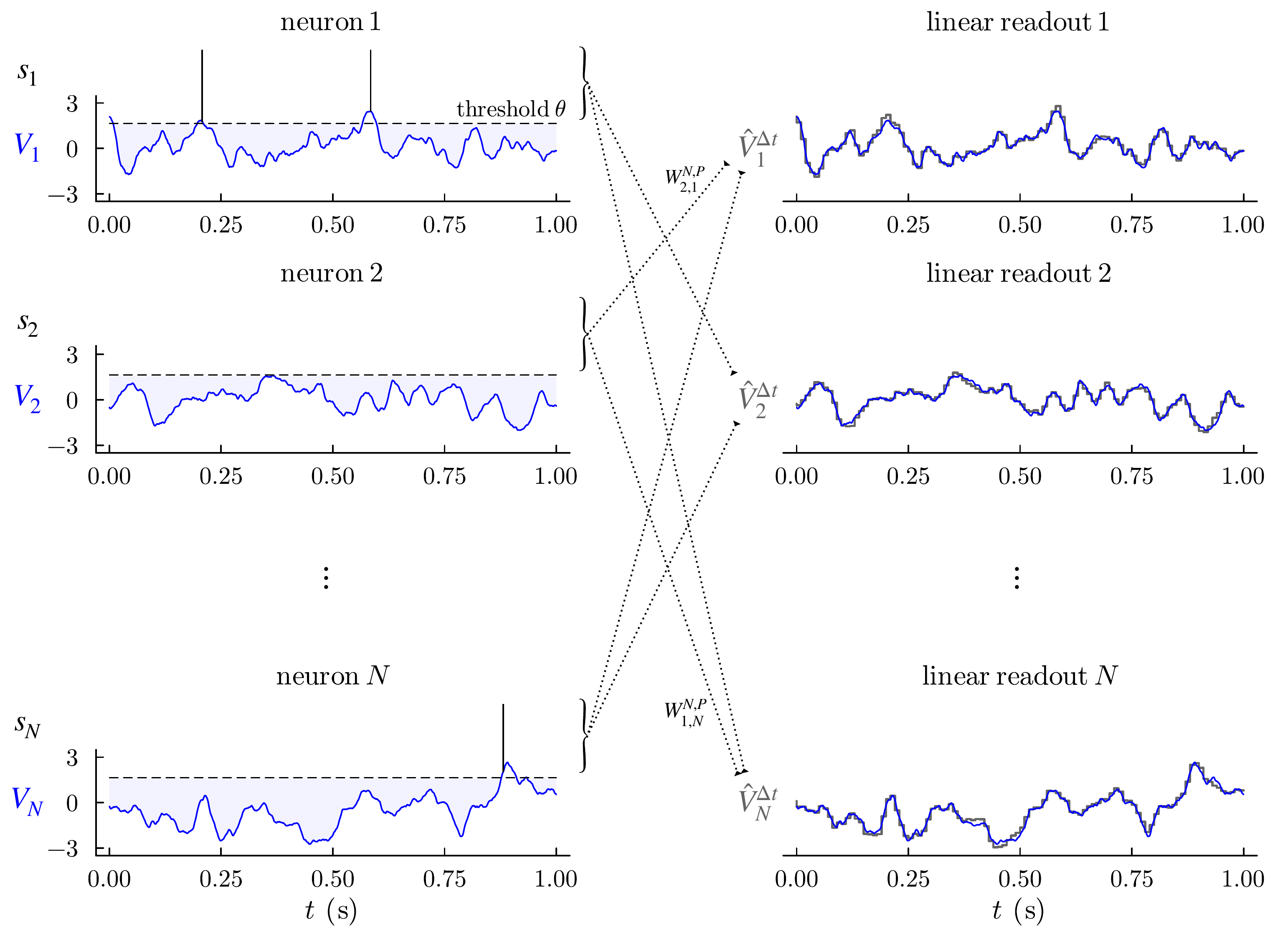}
    \caption{\label{fig:1}(\textbf{Left panels}) Presynaptic membrane potentials and spike trains of $N$ linear-nonlinear-Poisson neurons. The potentials $V_1(t), V_2(t), \dots, V_N(t)$ (blue lines).
    The instantaneous firing rate of a neuron depends nonlinearly on its potentials. Here, the nonlinear transfer function is the step function defined in \eqref{eq:threshold}: a neuron emits stochastic spikes at rate $\rho$ only when the potential is above the threshold $\theta$ (dashed line). Taking $\theta = 1.65$, the potentials spend approximately $95\%$ of the time below the threshold (light blue areas). Taking $\rho = 20\,\text{Hz}$, the average firing rate of each neuron is approximately $1\,\text{Hz}$; as a result, spikes $S_1(t), S_2(t), \dots, S_N(t)$ (vertical bars) are sparse. 
    At the single-neuron level, the spike train $S_i(t)$ contain little information about the potential $V_i(t)$.
    (\textbf{Right panels})  Comparison between the postsynaptic readouts and the presynaptic potentials. As predicted by the theorem, the readouts $\widehat{V}^{\Delta t}_1, \widehat{V}^{\Delta t}_2, \dots, \widehat{V}^{\Delta t}_N$ defined in \eqref{eq:readout} (gray step-lines) give near-exact approximations of the true potentials $V_1(t), V_2(t), \dots, V_N(t)$ (blue lines, same as on the left panels) when $N$ and $P$ are large and $P\ll N$. The small deviations between the gray and blue traces are due to the fact that, while $N$ and $P$ are large, they are finite; here, $N=10^6$, $P=100$, and $\Delta t = 2\,\text{ms}$. The readouts are weighted sums of the spike trains: the weights are schematically represented by dotted lines and two weights, $W_{2,1}^{N,P}$ and $W_{1,N}^{N,P}$ are labeled as a example. 
    At the network level, the spikes of a presynaptic population of neurons can perfectly transmit the presynaptic membrane potentials---including the subthreshold components (light blue area, left pannels)---to a postsynaptic population of linear readout neurons.
    The details of how the potentials $V_1(t), V_2(t), \dots, V_N(t)$ are generated in this figure are presented in Appendix~\ref{sec:details}.}
\end{figure*}

In our model, the presynaptic population of neurons consists of $N$ linear-nonlinear-Poisson neurons \cite{Chi01}. Each neuron $i=1,2,\dots,N$ has a time-varying membrane potential $V_i(t)$ (blue lines in left panels of Fig.~\ref{fig:1}) and it emits stochastic spikes with a time-varying firing rate, $\phi(V_i(t))$, which depends nonlinearly on the potential $V_i(t)$ through the non-negative transfer function $\phi$. For example, $\phi$ can be the step function
\begin{equation}\label{eq:threshold}
    \phi(V_i(t)) := \begin{cases}\rho, &\text{ if } V_i(t) \geq  \theta, \\ 0, &\text{ if } V_i(t) < \theta,\end{cases}
\end{equation}
with $\rho>0$. With this choice, a neuron can emit spikes only if its potential is above the threshold $\theta$ (dashed lines in left panels of Fig~\ref{fig:1}). In the linear-nonlinear-Poisson model, the spikes of a neuron are modeled as Dirac delta pulses that form the neuron's spike train $S_i(t) := \sum_{f}\delta(t -t_i^{(f)})$ (vertical bars in left panels of Fig.~\ref{fig:1}), where the spike times $t_i^{(1)}, t_i^{(2)}, \dots$ are generated from an inhomogeneous Poisson process with time-varying rate $\phi(V_i(t))$ \cite{Chi01,GerKis02}. For clarity, we mention that the variable $V_i(t)$ in this simple neuron model corresponds to the `coarse potential' of a real neuron, i.e., a low-pass filtered version of its membrane potential trace after the removal of the spikes \cite{CarFer00,Car04}.

Can the presynaptic neurons transmit their membrane potentials to a population of postsynaptic neurons through spikes? At first glance, when the transfer function $\phi$ has a threshold as in Eq.~\eqref{eq:threshold}, the subthreshold membrane potential fluctuations of a presynaptic neuron (light blue areas in the left panels of Fig.~\ref{fig:1}) cannot be transmitted to a postsynaptic neuron, since neuron~$i$ does not emit any spikes when $V_i(t)<\theta$. Contrary to this intuition, we show in the following that, for large enough populations, all the presynaptic potentials $V_1(t), V_2(t), \dots, V_N(t)$, including their subthreshold components, can be perfectly recovered by a postsynaptic population of \textit{linear} readout neurons. 

To model the postsynaptic neurons, it is formally convenient to consider spike counts over small time bins, since spikes are modeled as Dirac delta pulses. For any bin size $\Delta t > 0$ and for all time bin $[k\Delta t, (k+1)\Delta t[\,$ for $k\in\mathbb{N}$, we model postsynaptic neuron $i$ as the linear readout
\begin{equation}\label{eq:readout}
    \widehat{V}^{\Delta t}_{i,k} := \sum_{j=1}^N W_{i,j}\frac{1}{\Delta t}\int_{k\Delta t}^{(k+1)\Delta t}S_j(t)\rd t,
\end{equation}
where the $W_{i,j}$'s are the synaptic weights (schematically represented as dotted lines in Fig.~\ref{fig:1}). We stress that the readout $\widehat{V}^{\Delta t}_{i,k}$ only has access to the spikes emitted by presynaptic neurons (accolades in Fig.~\ref{fig:1}). We say that spikes can transmit neurons' membrane potentials if we can choose the weights in Eq.~\eqref{eq:readout} such that for all $i=1,2,\dots,N$, the postsynaptic readout $\widehat{V}^{\Delta t}_{i,k}$ perfectly recovers the presynaptic potential $V_i(t)$. The question is then: What weights would enable such perfect recovery? It turns out that if the presynaptic potentials are weakly correlated in a precise sense stated below, such weights are straightforwardly given by the covariance matrix of the presynaptic membrane potentials.

Here, we assume that the fluctuations of the potentials $V_i(t)$ are given by time-continuous, stationary stochastic processes. Moreover, we assume that, at any time $t$, the joint probability distribution of the potentials $V_1(t), V_2(t), \dots, V_N(t)$ follows a zero-mean, multivariate normal distribution $\mathcal{N}(\textbf{0},\mathbf{C}^{N,P})$ with covariance matrix given by
    \begin{equation}\label{eq:def_Sigma}
        \mathbf{C}^{N,P} := \frac{1}{P}\bxi\bxi^{\mathrm{T}},
    \end{equation}
where $\bxi$ is a $N \times P$ random matrix with $i.i.d.$ standard normal entries (i.e., $\mathbf{C}^{N,P}$ is sampled from the standard Wishart distribution $W_N(\frac{1}{P}I_P, P)$). When $P < N$, the parameter $P$ is the dimension of the random linear subspace spanned by the fluctuations of the potentials.
Under these assumptions, each potential $V_i(t)$ has mean zero and approximately unit variance when $P$ is large and the off-diagonal entries of the covariance matrix, i.e., the cross-correlations, are of order $1/\sqrt{P}$ for large $P$. When $P$ is large and $N$ is even larger, this construction of the covariance matrix leads to \textit{weak correlations} of the membrane potentials, namely, cross-correlations that are small but above chance level \cite{PouPet08} and greater than in the theoretical `asynchronous state' \cite{RenDel10}.

Assuming that the covariance matrix of the presynaptic potentials has the structure specified in Eq.~\eqref{eq:def_Sigma}, the synaptic weights that allow the postsynaptic neurons to recover the presynaptic potentials are simply given by a rescaled version of the covariance matrix,
\begin{equation}\label{eq:weights}
    W^{N,P}_{i,j}:=\begin{cases}m^\phi\frac{P}{N-1}C^{N,P}_{i,j}, &i\neq j,\\
    0, &i=j,
    \end{cases}
\end{equation}
where $m^\phi:= (\int z \phi(z)\frac{1}{\sqrt{2\pi}}e^{-z^2/2}\rd z)^{-1}$ is a constant that only depends on the transfer function $\phi$. Note that postsynaptic neuron $i$ does not need to receive any input from presynaptic neuron $i$ (see absence of dotted lines in Fig.~\ref{fig:1}). 

Substituting the synaptic weights proposed in Eq.~\eqref{eq:weights} into the definition of the readouts in Eq.~\eqref{eq:readout}, we quantify the distance between the postsynaptic readouts $\widehat{V}^{\Delta t}_{i,k}$ and the presynaptic potentials $V_i(t)$ using the mean squared error
\begin{equation*}
    E^{\Delta t}_{k}(N,P) := \frac{1}{N}\sum_{i=1}^N\left(\widehat{V}^{\Delta t}_{i,k} - \frac{1}{\Delta t}\int_{k\Delta t}^{(k+1)\Delta t}V_i(t)\rd t\right)^2.
\end{equation*}

\begin{theorem*}\label{theorem}
    Let $\phi:\R\to\R_+$ be a non-constant, monotonically increasing function with at most polynomial growth, i.e., there exists an exponent $\alpha\geq 0$ such that $\limsup_{v\to+\infty}\frac{\phi(v)}{|v|^\alpha}<+\infty$.
    Then, if the membrane potential fluctuations satisfy the assumptions described above, Eq.~\eqref{eq:def_Sigma}, and if $P$ grows with $N$ such that $P\to\infty$ and $P/N \to 0$ as $N\to\infty$, we have
    \begin{equation*}
        \E\left[E^{\Delta t}_k(N,P)\right] \xrightarrow[N\to\infty]{}0.
    \end{equation*}
\end{theorem*}
Informally, the theorem says that when the number of neurons $N$ is large enough and when $P$ (the linear dimensionality of the potential fluctuations) is large but small compared to $N$, the linear readouts $\widehat{V}^{\Delta t}_{i,k}$ give near-exact approximations of the true potentials $V_i(t)$. We emphasize that the theorem holds for a general class of transfer function $\phi$. Notably, the function $\phi$ does not need to be continuous, e.g., it can be a step function as in Eq.~\eqref{eq:threshold}. More generally, $\phi$ can be any rectified power function of the form $\phi(v) = \rho\times([v-\theta]_+)^\alpha$ with scale $\rho>0$, threshold $\theta \in \R$, and exponent $\alpha \geq 0$, as in \cite{Car04,MecRin02}. The proof of the theorem is presented in Appendices~\ref{sec:proof_theorem} and \ref{sec:proofs_lemmas}. It is inspired by results from high-dimensional probability \cite{Ver18} and involves concentration of measure arguments applied to networks of spiking neurons \cite{SchBre25}. Our result shows that from the perspective of a postsynaptic readout neuron, not only do the presynaptic spiking neurons behave like continuous rate units \cite{SchBre25}, but the nonlinearity $\phi$ itself---which can have a dramatic effect on a neuron's firing rate (see left panels in Fig.~\ref{fig:1})---becomes immaterial in the high-dimensional limit described in the theorem. 

The connectivity defined in Eq.~\eqref{eq:weights} is dense, with individual weights being of order $\sqrt{P}/N$ for large $N$ and $P$ (since cross-correlations are of order $1/\sqrt{P}$). In the high-dimensional limit described in the theorem, the scaling of the weights in $N$ falls strictly between $1/N$ and $1/\sqrt{N}$, an intermediate scaling regime that is consistent with the empirical $N^{-0.59}$ weight scaling measured in cortical neuron cultures \cite{BarRey16}. While cortical neuron cultures are densely connected (the number of input synapses of a neuron scales linearly with the total number of neurons $N$ \cite{BarRey16}), real brains are more sparsely connected, a pyramidal neuron having around $12,000$ input synapses in mice and $15,000$ input synapses in humans \cite{LooStr22}. Another aspect in which our synaptic weights Eq.~\eqref{eq:weights} are biologically simplistic is that they do not obey Dale's law, which would require all the output weights of a neuron to share the same sign. Whether or not the theorem can be adapted to synaptic weight matrices satisfying sparsity constraints or sign constraints is an open mathematical problem.

Our model assumed that the covariance of the membrane potentials follows a somewhat specific structure (see Eq.~\eqref{eq:def_Sigma}). While we do not know if this covariance structure is biologically realistic, we note that it leads to weak correlations between potentials, a hallmark of the `desynchronized state' of cortical activity observed when animals are directing their attention to sensory stimuli \cite{PouPet08,HarThi11}. Hence, our theorem provides a possible explanation for why the processing of sensory signals is improved in the desynchronized state \cite{HarThi11,McgDav15}, namely, weak correlations in the desynchronized state may be the reflection of a high-dimensional regime enabling network-level transmission of subthreshold information. This hypothesis leads to an experimental prediction: the membrane potentials of a large ensemble of neurons are more accurately recovered from linear readouts of their spikes in the desynchronized (higher dimensional) state than in the synchronized (lower dimensional) state. The rapid development of voltage imaging methods \cite{AdaKim19,VilCha19,FanKim23} may allow experimentalists to test this prediction in the near future.




\paragraph{Acknowledgments.} I thank Wulfram Gerstner for supporting this work and for comments on the manuscript. I also thank Alireza Modirshanechi and Maxwell Shinn for discussions and comments on the manuscript. This work was supported by the Swiss National Science Foundation (grant no. 200020\_207426 awarded to Wulfram Gerstner), a Royal Society Newton International Fellowship (NIF$\backslash$R1$\backslash$231927), and a Swiss National Science Foundation Postdoc.Mobility Fellowship (P500PB\_222150).

\newpage
\appendixpage
\appendix
\section{Proof of the theorem}\label{sec:proof_theorem}
To show that the expected mean squared error $\E\left[E^{\Delta t}(N,P)\right]$ between the potentials $\widehat{V}^{\Delta t}_i$ read out from the spikes, Eq.~\eqref{eq:readout} in the main text, and the true potential $V_i$ converges to $0$ as $N\to\infty$, if $P\to\infty$ and $P/N \to 0$ as $N\to\infty$, we will follow a kind of bias-variance decomposition and show that the bias of the estimator $\widehat{V}^{\Delta t}_1$ tends to $0$ as $P\to\infty$ and its variance also tends to $0$ as $N\to\infty$ if $P/N \to 0$.

Before undertaking the bias-variance decomposition, let us make some observations that directly follow from the definition of the model. Let $\{X_\mu\}_{\mu=1}^\infty$ be a sequence of \textit{i.i.d.} standard normal variables that are also independent of the $N \times P$ random matrix $\bxi$. By the definition of the membrane potential fluctuations $V_1(t), V_2(t), \dots, V_N(t)$, at any given time $t$, the $N$-dimensional random vector $(V_1(t), V_2(t), \dots, V_N(t))$, conditioned on $\bxi$, has the same law as
\begin{equation}\label{eq:latent}
    \left(\frac{1}{\sqrt{P}}\sum_{\mu=1}^{P}\xi_{1,\mu}X_\mu , \frac{1}{\sqrt{P}}\sum_{\mu=1}^{P}\xi_{2,\mu}X_\mu, \dots, \frac{1}{\sqrt{P}}\sum_{\mu=1}^{P}\xi_{N,\mu}X_\mu\right).
\end{equation}
The variables $X_1, X_2, \dots, X_P$ above can be interpreted as latent variables and the rows of the matrix $\bxi$ can be interpreted as random features. By the definition of the stochastic spike trains $S_1(t), S_2(t), \dots, S_N(t)$, conditioned on the potentials $V_1(t), V_2(t), \dots, V_N(t)$, the spike trains $S_1(t), S_2(t), \dots, S_N(t)$ are independent Poisson processes with time-varying intensities $\phi(V_1(t)),\phi(V_2(t)),\dots,\phi(V_N(t))$, respectively.

We can now begin the bias-variance decomposition. By the exchangeability of the neuron indices $i=1\dots,N$, 
\begin{align}
    \E\left[E^{\Delta t}_k(N,P)\right] &= \E\,\frac{1}{N}\sum_{i=1}^N\left(\widehat{V}^{\Delta t}_{i,k} - \frac{1}{\Delta t}\int_{k\Delta t}^{(k+1)\Delta t}V_i(t)\rd t\right)^2 \nonumber\\
    &= \E\left(\widehat{V}^{\Delta t}_{1,k} - \frac{1}{\Delta t}\int_{k\Delta t}^{(k+1)\Delta t}V_1(t)\rd t\right)^2 \nonumber\\
    &= \E\left(m^\phi\frac{P}{N-1}\sum_{i=2}^N C^{N,P}_{1,i}\frac{1}{\Delta t}\int_{k\Delta t}^{(k+1)\Delta t}S_i(t)\rd t - \frac{1}{\Delta t}\int_{k\Delta t}^{(k+1)\Delta t}V_1(t)\rd t\right)^2. \label{eq:neuron_1}
\end{align}
For each $i=1\dots,N$, let us introduce the compensated jump process $t\mapsto \int_{t_0}^t [S_i(u) - \phi(V_i(u))]\rd u$, which satisfies
\begin{equation}\label{eq:compensated}
    \E\int_{t_0}^t \left[S_i(u) - \phi(V_i(u))\right]\rd u  = 0, \quad \forall t\geq t_0.
\end{equation}
Inserting these compensated processes in \eqref{eq:neuron_1} and using \eqref{eq:compensated},
\begin{multline}\label{eq:first-BV}
    \E\left[E^{\Delta t}_k(N,P)\right] = \E\left(m^\phi\frac{P}{N-1}\sum_{i=2}^N C^{N,P}_{1,i}\frac{1}{\Delta t}\int_{k\Delta t}^{(k+1)\Delta t}\left[S_i(t) - \phi(V_i(t))\right]\rd t \right)^2 \\
     + \E\left(m^\phi\frac{P}{N-1}\sum_{i=2}^N C^{N,P}_{1,i}\frac{1}{\Delta t}\int_{k\Delta t}^{(k+1)\Delta t}\phi(V_i(t))\rd t - \frac{1}{\Delta t}\int_{k\Delta t}^{(k+1)\Delta t}V_1(t)\rd t\right)^2.
\end{multline}
The first term on the right-hand side of \eqref{eq:first-BV}, 
\begin{equation*}
   \E\left(m^\phi\frac{P}{N-1}\sum_{i=2}^N C^{N,P}_{1,i}\frac{1}{\Delta t}\int_{k\Delta t}^{(k+1)\Delta t}\left[S_i(t) - \phi(V_i(t))\right]\rd t \right)^2 =: \sigma^2_{\text{Poisson-noise}}(N,P),
\end{equation*}
has to be interpreted as the variance due to Poisson spike noise. 

\subsection*{The variance due to spike noise, $\sigma^2_{\text{Poisson-noise}}(N,P)$, vanishes as $N\to\infty$ if $P/N \to 0$:}

Since the spike trains $S_1(t),S_2(t),\dots,S_N(t)$ are conditionally independent Poisson processes given the potentials $V_1(t), V_2(t), \dots, V_N(t)$, we have
\begin{align*}
    \sigma^2_{\text{Poisson-noise}}(N,P) &= \E\left(m^\phi\frac{P}{N-1}\sum_{i=2}^N C^{N,P}_{1,i}\frac{1}{\Delta t}\int_{k\Delta t}^{(k+1)\Delta t}\left[S_i(t) - \phi(V_i(t))\right]\rd t \right)^2 \\
    &= \left(\frac{m^\phi}{\Delta t}\right)^2 \frac{P^2}{N-1}\E\left[(C^{N,P}_{1,2 })^2 \left(\int_{k\Delta t}^{(k+1)\Delta t}\left[s_2(t) - \phi(V_2(t))\right]\rd t\right)^2\right].
\end{align*}
By Itô's isometry for compensated jump processes \cite[Lemma~4.2.2 p.~197]{App09}, 
\begin{equation*}
    \E\left[\left(\int_{k\Delta t}^{(k+1)\Delta t}\left[s_2(t) - \phi(V_2(t))\right]\rd t\right)^2 \Bigg| \,V_2\right] = \int_{k\Delta t}^{(k+1)\Delta t}\phi(V_2(t))^2\rd t.
\end{equation*}
Then, using the representation~\eqref{eq:latent} for the stationary distribution of the potentials, we get
\begin{align*}
    \sigma^2_{\text{Poisson-noise}}(N,P) &= \frac{(m^\phi)^2}{\Delta t}\frac{1}{N-1}\E\left[\left(\sum_{\mu=1}^P\xi_{1,\mu}\xi_{2,\mu}\right)^2 \phi\left(\frac{1}{\sqrt{P}}\sum_{\mu=1}^P\xi_{2,\mu}X_\mu\right)^2\right] \\
    &= \frac{(m^\phi)^2}{\Delta t}\frac{1}{N-1}\sum_{\mu=1}^P\underbrace{\E\left[\xi_{1,\mu}^2\right]}_{=1}\E\left[\xi_{2,\mu}^2 \phi\left(\frac{1}{\sqrt{P}}\sum_{\mu=1}^P\xi_{2,\mu}X_\mu\right)^2\right] \\
    &= \frac{(m^\phi)^2}{\Delta t}\frac{1}{N-1}\E\left[\sum_{\mu=1}^P\xi_{2,\mu}^2 \phi\left(\frac{1}{\sqrt{P}}\sum_{\mu=1}^P\xi_{2,\mu}X_\mu\right)^2\right].
\end{align*}

To bound the expectation $\E\left[\sum_{\mu=1}^P\xi_{2,\mu}^2 \phi\left(\frac{1}{\sqrt{P}}\sum_{\mu=1}^P\xi_{2,\mu}X_\mu\right)^2\right]$ above, we will use two lemmas, whose proofs are postponed to Appendix~\ref{sec:proofs_lemmas}. These lemmas will also be used later in the proof.



\begin{lemma}\label{lemma:expectation} For any $P\geq 1$, let $\zeta_1, \zeta_2, \dots, \zeta_P$ be \textit{i.i.d.} standard normal variables and let us write $\bzeta := (\zeta_1, \zeta_2, \dots, \zeta_P)$ the corresponding random vector. Then, for any given vector $\x  := (x_1, x_2, \dots, x_P)\in \R^P$, 
\begin{equation}\label{eq:expectation}
    \E\left[\bzeta \phi \left(\frac{1}{\sqrt{P}}\bzeta^{\mathrm{T}}\x \right)\bigg | \,\x \right] = \frac{\x}{\|\x\|}\int z\phi\left(\frac{\|\x\|}{\sqrt{P}}z\right)\cD z
\end{equation}
and
\begin{equation}\label{eq:second_moment}
    \E\left[\|\bzeta\|^2 \phi \left(\frac{1}{\sqrt{P}}\bzeta^{\mathrm{T}}\x \right)^2\bigg | \,\x \right] = \int z^2\phi\left(\frac{\|\x\|}{\sqrt{P}}z\right)^2\cD z + (P-1)\int \phi\left(\frac{\|\x\|}{\sqrt{P}}z\right)^2\cD z,
\end{equation}
where $\mathcal{D}z := \frac{1}{\sqrt{2\pi}}e^{-z^2/2}\rd z$ denotes the standard Gaussian measure.
\end{lemma}
We mention that a result similar to \eqref{eq:expectation} in Lemma~\ref{lemma:expectation} is used, but under a different form, in the theory of low-rank recurrent neural networks; see, e.g., \cite{BeiDub21}.

By \eqref{eq:second_moment} in Lemma~\ref{lemma:expectation}, we get
\begin{equation*}
    \E\left[\sum_{\mu=1}^P\xi_{2,\mu}^2 \phi\left(\frac{1}{\sqrt{P}}\sum_{\mu=1}^P\xi_{2,\mu}X_\mu\right)^2\right] = \E\int z^2\phi\left(\frac{\|\X_{1:P}\|}{\sqrt{P}}z\right)^2\cD z + (P-1)\;\E\int \phi\left(\frac{\|\X_{1:P}\|}{\sqrt{P}}z\right)^2\cD z,
\end{equation*}
where $\X_{1:P} := (X_1, X_2, \dots, X_P)$. Note that since $X_1, X_2, \dots, X_P$ are \textit{i.i.d.} standard normal variables, $\|\X_{1:P}\|$ follows a Chi distribution with $P$ degrees of freedom.
\begin{lemma}\label{lemma:bounds}
Let $\chi(P)$ denote the Chi distribution with $P$ degrees of freedom. Then, for all $\beta\geq 0$ and $\gamma\geq 0$,
\begin{equation*}
    Q^{\beta,\gamma}:=\sup_{P\geq 1}\E_{Y\sim \chi(P)}\int |z|^\beta\phi\left(Yz/\sqrt{P}\right)^\gamma\cD z<+\infty.
\end{equation*}
\end{lemma}

By Lemma~\ref{lemma:bounds}, 
\begin{equation*}
\E\left[\sum_{\mu=1}^P\xi_{2,\mu}^2 \phi\left(\frac{1}{\sqrt{P}}\sum_{\mu=1}^P\xi_{2,\mu}X_\mu\right)^2\right] \leq Q^{2,2} + (P-1)\,Q^{0,2}.
\end{equation*}
Hence,
\begin{equation*}
    \sigma^2_{\text{Poisson-noise}}(N,P) \leq \frac{(m^\phi)^2}{\Delta t}\frac{Q^{2,2} + (P-1)\,Q^{0,2}}{N-1},
\end{equation*}
which tends to $0$ as $N\to\infty$ if $P/N\to 0$.

We now proceed with the bias-variance decomposition by turning to the second term on the right-hand side of \eqref{eq:first-BV}. By the measure-theoretic form of Jensen's inequality \cite[Theorem~3.3 p.~62]{Rud87},
\begin{multline*}
    \E\left(m^\phi\frac{P}{N-1}\sum_{i=2}^N C^{N,P}_{1,i}\frac{1}{\Delta t}\int_{k\Delta t}^{(k+1)\Delta t}\phi(V_i(t))\rd t - \frac{1}{\Delta t}\int_{k\Delta t}^{(k+1)\Delta t}V_1(t)\rd t\right)^2 \\
    \leq \E \frac{1}{\Delta t}\int_{k\Delta t}^{(k+1)\Delta t}\left(m^\phi\frac{P}{N-1}\sum_{i=2}^N C^{N,P}_{1,i} \phi(V_i(t)) - V_1(t)\right)^2\rd t.
\end{multline*}
Then, using the representation~\eqref{eq:latent} for the stationary distribution of the potentials, we get
\begin{multline*}
    \E \frac{1}{\Delta t}\int_{k\Delta t}^{(k+1)\Delta t}\left(m^\phi\frac{P}{N-1}\sum_{i=2}^N C^{N,P}_{1,i} \phi(V_i(t)) - V_1(t)\right)^2\rd t \\
    = \E\left(m^\phi\frac{1}{N-1}\sum_{i=2}^N \sum_{\mu=1}^P\xi_{1,\mu}\xi_{i,\mu} \phi\left(\frac{1}{\sqrt{P}}\sum_{\nu=1}^P \xi_{i,\nu}X_\nu\right) - \frac{1}{\sqrt{P}}\sum_{\nu=1}^P \xi_{1,\nu}X_\nu\right)^2
\end{multline*}
Since the entries of $\bxi$ are \textit{i.i.d.} standard normal variables, 
\begin{align*}
    &\E\left(\frac{m^\phi}{N-1}\sum_{i=2}^N \sum_{\mu=1}^{P}\xi_{1,\mu}\xi_{i,\mu}\phi\left(\frac{1}{\sqrt{P}}\sum_{\nu=1}^P\xi_{i,\nu}X_\nu\right) - \frac{1}{\sqrt{P}}\sum_{\mu=1}^P\xi_{1,\mu}X_\mu\right)^2\\
    &\qquad=\E\left(\frac{1}{\sqrt{P}}\sum_{\mu=1}^{P}\xi_{1,\mu}\left(\frac{m^\phi \sqrt{P}}{N-1}\sum_{i=2}^N \xi_{i,\mu}\phi\left(\frac{1}{\sqrt{P}}\sum_{\nu=1}^P\xi_{i,\nu}X_\nu\right) -X_\mu\right)\right)^2 \\
    &\qquad=\frac{1}{P}\sum_{\mu=1}^{P}\E\left(\frac{m^\phi \sqrt{P}}{N-1}\sum_{i=2}^N \xi_{i,\mu}\phi\left(\frac{1}{\sqrt{P}}\sum_{\nu=1}^P\xi_{i,\nu}X_\nu\right) - X_\mu\right)^2.
\end{align*}

We can now make the final bias-variance decomposition. Let $\{\zeta_k\}_{k=1}^\infty$ a sequence of $i.i.d.$ standard normal random variables that are also independent of $\{X_k\}_{k=1}^\infty$.
\begin{multline}
    \frac{1}{P}\sum_{\mu=1}^{P}\E\left[\left(\frac{m^\phi \sqrt{P}}{N-1}\sum_{i=2}^N \xi_{i,\mu}\phi\left(\frac{1}{\sqrt{P}}\sum_{\nu=1}^P\xi_{i,\nu}X_\nu\right) - X_\mu\right)^2\right] \\
    = \underbrace{(m^\phi)^2\sum_{\mu=1}^{P}\E\left(\frac{1}{N-1}\sum_{i=2}^N \xi_{i,\mu}\phi\left(\frac{1}{\sqrt{P}}\sum_{\nu=1}^P\xi_{i,\nu}X_\nu\right) - \E\left[\zeta_\mu \phi\left(\frac{1}{\sqrt{P}}\sum_{\nu=1}^P \zeta_\nu X_\nu\right)\Bigg|\,\X\right]\right)^2}_{=:\sigma^2_{\bxi-\mathrm{noise}}(N,P)} \\
    + \underbrace{\frac{1}{P}\sum_{\mu=1}^P\E\left(m^\phi\sqrt{P}\, \E\left[\zeta_\mu \phi\left(\frac{1}{\sqrt{P}}\sum_{\nu=1}^P \zeta_\nu X_\nu\right)\Bigg|\,\X\right] - X_\mu\right)^2}_{=:b(P)}. \label{eq:second_BV}
\end{multline}
The first term on the right-hand side of \eqref{eq:second_BV}, $\sigma^2_{\bxi-\mathrm{noise}}(N,P)$, has to be interpreted as the variance due to the random matrix $\bxi$; the second term, $b(P),$ is the expected squared bias of linear readout Eq.~\eqref{eq:readout} in the main text.

\subsection*{The variance due to the random matrix $\bxi$, $\sigma^2_{\bxi-\mathrm{noise}}(N,P)$, vanishes as $N\to\infty$ if $P/N\to 0$:} 
Expanding the square in the definition of $\sigma^2_{\bxi-\mathrm{noise}}(N,P)$, we get
\begin{multline*}
    (m^\phi)^2\sum_{\mu=1}^{P}\E\left(\frac{1}{N-1}\sum_{i=2}^N \xi_{i,\mu}\phi\left(\frac{1}{\sqrt{P}}\sum_{\nu=1}^P\xi_{i,\nu}X_\nu\right) - \E\left[\zeta_\mu \phi\left(\frac{1}{\sqrt{P}}\sum_{\nu=1}^P \zeta_\nu X_\nu\right)\Bigg|\,\X\right]\right)^2\\
    = (m^\phi)^2\sum_{\mu=1}^{P}\frac{1}{(N-1)^2}\sum_{i=2}^N\E \left(\xi_{i,\mu}\phi\left(\frac{1}{\sqrt{P}}\sum_{\nu=1}^P\xi_{i,\nu}X_\nu\right) - \E\left[\zeta_\mu \phi\left(\frac{1}{\sqrt{P}}\sum_{\nu=1}^P \zeta_\nu X_\nu\right)\Bigg|\,\X\right]\right)^2.
\end{multline*}
Then, using
\begin{equation*}
    \E\left(\xi_{i,\mu}\phi\left(\frac{1}{\sqrt{P}}\sum_{\nu=1}^P\xi_{i,\nu}X_\nu\right) - \E\left[\zeta_\mu \phi\left(\frac{1}{\sqrt{P}}\sum_{\nu=1}^P \zeta_\nu X_\nu\right)\Bigg|\,\X\right]\right)^2 \leq \E\left[\zeta_{\mu}^2\phi\left(\frac{1}{\sqrt{P}}\sum_{\nu=1}^P\zeta_{\nu}X_\nu\right)^2\right],
\end{equation*}
we get
\begin{align*}
    \sigma^2_{\bxi-\mathrm{noise}}(N,P) 
    &\leq (m^\phi)^2\sum_{\mu=1}^{P}\frac{1}{N-1}\E\left[\zeta_{\mu}^2\phi\left(\frac{1}{\sqrt{P}}\sum_{\nu=1}^P\zeta_{\nu}X_\nu\right)^2\right] \\
    &= (m^\phi)^2\frac{1}{N-1}\E\left[\|\bzeta\|^2\phi\left(\frac{1}{\sqrt{P}}\sum_{\mu=1}^P\zeta_{\mu}X_\mu\right)^2\right].
\end{align*}
Using \eqref{eq:second_moment} in Lemma~\ref{lemma:expectation} and Lemma~\ref{lemma:bounds}, we find the bound
\begin{equation*}
    \sigma^2_{\bxi-\mathrm{noise}}(N,P) 
    \leq (m^\phi)^2\frac{Q^{2,2} + (P-1)\,Q^{0,2}}{N-1},
\end{equation*}
which tends to $0$ as $N\to\infty$ if $P/N \to 0$.

\subsection*{The expected squared bias, $b(P)$, vanishes as $N\to\infty$ if $P\to\infty$:} 
Let us write $\X_{1:P} := (X_1, X_2, \dots, X_P)$, for any $P\geq 1$. Using \eqref{eq:expectation} in Lemma~\ref{lemma:expectation}, we have
\begin{align*}
    b(P) &= \frac{1}{P}\sum_{\mu=1}^P\E\left(\frac{\sqrt{P}}{\|\X_{1:P}\|}m^\phi\int z\phi\left(\frac{\|\X_{1:P}\|}{\sqrt{P}}z\right)\cD z\,X_\mu - X_\mu\right)^2 \\
    &= \frac{1}{P}\sum_{\mu=1}^P\E\left[X_\mu^2\left(\frac{\sqrt{P}}{\|\X_{1:P}\|}m^\phi\int z\phi\left(\frac{\|\X_{1:P}\|}{\sqrt{P}}z\right)\cD z - 1\right)^2\right] \\
    &= \E\frac{1}{P}\sum_{\mu=1}^P X_\mu^2\left(\frac{\sqrt{P}}{\|\X_{1:P}\|}m^\phi\int z\phi\left(\frac{\|\X_{1:P}\|}{\sqrt{P}}z\right)\cD z - 1\right)^2.
\end{align*}
Noticing that $\frac{1}{P}\sum_{\mu=1}^P X_\mu^2 = \frac{\|\X_{1:P}\|^2}{P}$, we get
\begin{align*}
    b(P) &= \E\left[\frac{\|\X_{1:P}\|^2}{P}\left(\frac{\sqrt{P}}{\|\X_{1:P}\|}m^\phi\int z\phi\left(\frac{\|\X_{1:P}\|}{\sqrt{P}}z\right)\cD z - 1\right)^2\right] \\
    &= \E\left(m^\phi\int z\phi\left(\frac{\|\X_{1:P}\|}{\sqrt{P}}z\right)\cD z - \frac{\|\X_{1:P}\|}{\sqrt{P}}\right)^2.
\end{align*}
We state a last lemma, whose proof is presented in Appendix~\ref{sec:proof_lemma_3}.
\begin{lemma}\label{lemma:bias}
    \begin{equation}\label{eq:bias}
        \E\left(m^\phi\int z\phi\left(\frac{\|\X_{1:P}\|}{\sqrt{P}}z\right)\cD z - \frac{\|\X_{1:P}\|}{\sqrt{P}}\right)^2 \xrightarrow[P\to\infty]{} 0.
    \end{equation}
\end{lemma}
Since $P\to \infty$ as $N\to\infty$, we obtain, by Lemma~\ref{lemma:bias}, that the expected squared bias $b(P)$ tends to $0$ as $N\to\infty$.


In summary, we have established that the expected mean squared error of the linear readout, Eq.~\eqref{eq:readout} in the main text, is bounded by the sum of three terms:
\begin{equation*}
    \E\left[E^{\Delta t}_k(N,P)\right] \leq \sigma^2_{\text{Poisson-noise}}(N,P) + \sigma^2_{\bxi-\mathrm{noise}}(N,P) + b(P).
\end{equation*}
The two variance terms $\sigma^2_{\text{Poisson-noise}}(N,P)$ and $\sigma^2_{\bxi-\mathrm{noise}}(N,P)$ tend to $0$ as $N\to\infty$ if $N/P\to\infty$; the bias term $b(P)$ tends to $0$ as $N\to\infty$ is $P\to\infty$. Therefore, we have proved that the expected mean squared error of the linear readout vanishes as $N\to\infty$ if $P\to\infty$ and $P/N\to\infty$.

\section{Proofs of the Lemmas}\label{sec:proofs_lemmas}
\subsection{Proof of Lemma~\ref{lemma:expectation}}\label{sec:proof_lemma_1}
Consider the first component of the left hand side of Eq.~\eqref{eq:expectation},
\begin{equation*}
    \E\left[\zeta_1 \phi \left(\frac{1}{\sqrt{P}}\bzeta^{\mathrm{T}}\x \right)\bigg | \,\x \right] = \int\int \dots \int z_1 \phi\left(\frac{1}{\sqrt{P}}\sum_{\mu=1}^P z_\mu x_\mu\right)\cD z_1\cD z_2 \dots \cD z_P,
\end{equation*}
and apply the change of variable
\begin{equation}\label{eq:M}
    \begin{pmatrix}z_1 \\ z_2 \\ \vdots \\ z_P\end{pmatrix} = \underbrace{\begin{pmatrix}\frac{x_1}{\|\x\|} & a_{1,2} & \dots & a_{1,P}\\
    \frac{x_2}{\|\x\|} & a_{2,2} & \dots & a_{2,P}\\
    \vdots & \vdots & \ddots & \vdots \\
    \frac{x_P}{\|\x\|} & a_{P,2} & \dots & a_{P,P}
    \end{pmatrix}}_{=:\M}\begin{pmatrix}y_1 \\ y_2 \\\vdots \\ y_P\end{pmatrix},
\end{equation}
where $\M$ is an orthonormal matrix (i.e. $\M^\mathrm{T} \M = \M \M^\mathrm{T} = \mathbf{I}_P$). Notice that since $\M$ is orthonormal, $\sum_{\mu=1}^P x_\mu z_\mu = \|\x\|y_1$ and $|\mathrm{det}(\M)|=1$; hence, applying the change of variable~\eqref{eq:M},
\begin{align*}
    \int \int \dots \int z_1 \phi&\left(\frac{1}{\sqrt{P}}\sum_{\mu=1}^P z_\mu x_\mu\right)\cD z_1\cD z_2 \dots \cD z_P \\
    &= \int \int \dots \int \left(\frac{x_1}{\|\x\|}y_1 + \sum_{\mu=2}^P a_{1,\mu}y_\mu\right)\phi\left(\frac{\|\x\|}{\sqrt{P}}y_1\right)\cD y_1\cD y_2 \dots \cD y_P \\
    &= \frac{x_1}{\|\x\|}\int y_1\phi\left(\frac{\|\x\|}{\sqrt{P}}y_1\right)\cD y_1 + \sum_{\mu=2}^P a_{1,\mu}\int \phi\left(\frac{\|\x\|}{\sqrt{P}}y_1\right)\cD y_1 \int y_\mu \cD y_\mu.
\end{align*}
But since $\int y_\mu \cD y_\mu =0$, we get
\begin{equation*}
    \int \int \dots \int z_1 \phi\left(\frac{1}{\sqrt{P}}\sum_{\mu=1}^P z_\mu x_\mu\right)\cD z_1\cD z_2 \dots \cD z_P = \frac{x_1}{\|\x\|}\int y_1\phi\left(\frac{\|\x\|}{\sqrt{P}}y_1\right)\cD y_1. 
\end{equation*}
Since this is true for all components $\mu=1, \dots, P$, we obtain
\begin{equation*}
    \E\left[\bzeta \phi \left(\frac{1}{\sqrt{P}}\bzeta^{\mathrm{T}}\x \right)\bigg | \,\x \right] = \frac{\x}{\|\x\|}\int z\phi\left(\frac{\|\x\|}{\sqrt{P}}z\right)\cD z,
\end{equation*}
which concludes the proof of \eqref{eq:expectation}. The proof of \eqref{eq:second_moment} uses the same change of variable. 
\begin{align*}
    \E\left[\zeta_1^2 \phi \left(\frac{1}{\sqrt{P}}\bzeta^{\mathrm{T}}\x \right)^2\bigg | \,\x \right] &= \int\int \dots \int z_1^2 \phi\left(\frac{1}{\sqrt{P}}\sum_{\mu=1}^P z_\mu x_\mu\right)^2\cD z_1\cD z_2 \dots \cD z_P \\
    &= \int \int \dots \int \left(\frac{x_1}{\|\x\|}y_1 + \sum_{\mu=2}^P a_{1,\mu}y_\mu\right)^2\phi\left(\frac{\|\x\|}{\sqrt{P}}y_1\right)^2\cD y_1\cD y_2 \dots \cD y_P.
\end{align*}
Expanding the square and using that $\int y_\mu \cD y_\mu =0$ and $\int y_\mu^2 \cD y_\mu =1$, we get
\begin{equation*}
    \E\left[\zeta_1^2 \phi \left(\frac{1}{\sqrt{P}}\bzeta^{\mathrm{T}}\x \right)^2\bigg | \,\x \right] = \frac{x_1^2}{\|\x\|^2}\int y_1^2\phi\left(\frac{\|\x\|}{\sqrt{P}}y_1\right)^2\cD y_1 + \sum_{\mu=2}^P a_{1,\mu}^2\int \phi\left(\frac{\|\x\|}{\sqrt{P}}y_1\right)^2\cD y_1.
\end{equation*}
Since this is true for all components $\mu=1, \dots, P$, we get
\begin{align*}
    \E\left[\|\bzeta\|^2 \phi \left(\frac{1}{\sqrt{P}}\bzeta^{\mathrm{T}}\x \right)^2\bigg | \,\x \right] &= \sum_{\mu=1}^P\E\left[\zeta_1^2 \phi \left(\frac{1}{\sqrt{P}}\bzeta^{\mathrm{T}}\x \right)^2\bigg | \,\x \right]\\
    &= \sum_{\mu=1}^P \frac{x_\mu^2}{\|\x\|^2}\int y_1^2\phi\left(\frac{\|\x\|}{\sqrt{P}}y_1\right)^2\cD y_1 + \sum_{\mu=1}^P\sum_{\nu=2}^P a_{\mu,\nu}^2 \int\phi\left(\frac{\|\x\|}{\sqrt{P}}y_1\right)^2\cD y_1.
\end{align*}
Finally, we recall that $\sum_{\mu=1}^P \frac{x_\mu^2}{\|\x\|^2} = 1$ and $\sum_{\mu=1}^P a_{\mu,\nu}^2=1$, for all $\nu \geq 2$, since the matrix $\M$ \eqref{eq:M} is orthonormal, and find
\begin{equation*}
    \E\left[\|\bzeta\|^2 \phi \left(\frac{1}{\sqrt{P}}\bzeta^{\mathrm{T}}\x \right)^2\bigg | \,\x \right] = \int y_1^2\phi\left(\frac{\|\x\|}{\sqrt{P}}y_1\right)^2\cD y_1 + (P-1)\int\phi\left(\frac{\|\x\|}{\sqrt{P}}y_1\right)^2\cD y_1,
\end{equation*}
which concludes the proof of \eqref{eq:second_moment}.

\subsection{Proof of Lemma~\ref{lemma:bounds}}\label{sec:proof_lemma_2}
We recall that the function $\phi:\R\to\R_+$ is assumed to be monotonically increasing and to have polynomial growth at most.
By definition, there exist $v_0>0$, $C>0$, and $\alpha\geq 0$ such that $\phi(v) \leq Cv^\alpha$, for all $v\geq v_0$. Therefore, 
\begin{equation*}
    \phi(v)\leq Cv_0^\alpha + C|v|^\alpha, \quad \forall v\in\R.
\end{equation*}
Using the inequality above, we have, for any $P\geq 1$, 
\begin{equation*}
    \E_{Y\sim \chi(P)}\int |z|^\beta\phi\left(Yz/\sqrt{P}\right)^\gamma\cD z
     \leq  \left(\int|z|^\beta \cD z\right) C^\gamma v_0^{\alpha\gamma} + \E_{Y\sim \chi(P)}\int |z|^\beta C^\gamma \big|Yz/\sqrt{P}\big|^{\alpha\gamma}\cD z
\end{equation*}
In the inequality above, only the second term on the right hand side depends on $P$. This term does not diverge with $P$. Indeed,
\begin{equation*}
    \E_{Y\sim \chi(P)}\int |z|^\beta C^\gamma \big|Yz/\sqrt{P}\big|^{\alpha\gamma}\cD z = C^\gamma \left(\int |z|^{\beta+\alpha\gamma} \cD z\right) P^{-\alpha\gamma/2}\,\E_{Y\sim \chi(P)}\left[Y^{\alpha\gamma}\right].
\end{equation*}
The moments of the Chi distribution are finite and can be computed:
\begin{equation*}
    \E_{Y\sim \chi(P)}\left[Y^{\alpha\gamma}\right] = 2^{\alpha\gamma/2}\frac{\Gamma((\alpha\gamma + P)/2)}{\Gamma(P/2)},
\end{equation*}
where $\Gamma$ denotes the gamma function. By a property of the gamma function, he have the limit
\begin{equation*}
    \frac{\Gamma((\alpha\gamma + P)/2)}{\Gamma(P/2)(P/2)^{\alpha\gamma/2}} \xrightarrow[P\to\infty]{}1.
\end{equation*}
Hence, we conclude that
\begin{equation*}
    \sup_{P\geq 1}\E_{Y\sim \chi(P)}\int |z|^\beta\phi\left(Yz/\sqrt{P}\right)^\gamma\cD z <+\infty.
\end{equation*}
\subsection{Proof of Lemma~\ref{lemma:bias}}\label{sec:proof_lemma_3}
By the strong law of large numbers, 
\begin{equation*}
    \frac{\|\X_{1:P}\|^2}{P} = \frac{1}{P}\sum_{\mu=1}^P X_\mu^2\xrightarrow[P\to\infty]{a.s.} 1.
\end{equation*}
Since the function $\phi$ is monotonically increasing, it is almost everywhere continuous, and it is easy to verify that the function
\begin{equation*}
    G(x) := m^\phi\int z\phi\left(\sqrt{x}z\right)\cD z - \sqrt{x}
\end{equation*}
is continuous on $\R$. Moreover, by the definition of $m^\phi$, i.e. $m^\phi := (\int z \phi(z)\cD z)^{-1}$, we have $G(1) = 0$.
Then, we can apply the continuous mapping theorem (see \cite[Theorem~10.1 p.~244]{Gut06}) to obtain the almost sure convergence
\begin{equation*}
    \left(m^\phi\int z\phi\left(\frac{\|\X_{1:P}\|}{\sqrt{P}}z\right)\cD z - \frac{\|\X_{1:P}\|}{\sqrt{P}}\right)\xrightarrow[P\to\infty]{a.s.} 0.
\end{equation*}
To turn almost sure convergence above into a convergence in $L^2$ Eq.~\eqref{eq:bias}, we can use Vitali convergence theorem (see \cite[Theorem~5.2 p.~218]{Gut06}) by showing that the sequence of random variables
\begin{equation*}
    K_P := \left(m^\phi\int z\phi\left(\frac{\|\X_{1:P}\|}{\sqrt{P}}z\right)\cD z - \frac{\|\X_{1:P}\|}{\sqrt{P}}\right)^2, \quad \forall P\geq 1,
\end{equation*}
is \textit{uniformly integrable}. To show uniform integrability, we consider the sequence of dominating random variables $\{L_P\}_{P\geq 1}$,
\begin{equation*}
    K_P \leq \left(m^\phi\int z\phi\left(\frac{\|\X_{1:P}\|}{\sqrt{P}}z\right)\cD z\right)^2 + \frac{\|\X_{1:P}\|^2}{P}=: L_P.
\end{equation*}
Noticing that $\|\X_{1:P}\|$ and $\|\X_{1:P}\|^2$ follow a Chi and a Chi-squared distribution with $P$ degrees of freedom, respectively, and using Lemma~\ref{lemma:bounds}, we can easily verify that
\begin{equation*}
    \sup_{P\geq 1}\E\big[L_P^2\big] < +\infty.
\end{equation*}
Hence, we have that
\begin{equation*}
    \sup_{P\geq 1}\E\big[K_P^2\big] \leq \sup_{P\geq 1}\E\big[L_P^2\big] <+\infty,
\end{equation*}
which implies that the sequence of random variables $\{K_P\}_{P\geq 1}$ is uniformly integrable (see \cite[Theorem~4.2 p.~215]{Gut06}) and concludes the proof. 

\section{Generation of the membrane potentials in Fig.~\ref{fig:1}}\label{sec:details}
The membrane potentials $V_1(t), V_2(t), \dots, V_N(t)$ in Fig.~\ref{fig:1} are stationary Gaussian processes obtained through the numerical integration of the following system of stochastic differential equations: for all $i=1,\dots,N$,
\begin{align*}
    \tau\frac{\rd V_i(t)}{\rd t} &= -V_i(t) + A_i(t), \\
    \tau \rd A_i(t) &= -A_i(t)\rd t + \frac{2\sqrt{\tau}}{\sqrt{P}}\sum_{\mu=1}^P\xi_{i,\mu}\rd B_\mu(t),
\end{align*}
where $B_1(t), B_2(t), \dots, B_P(t)$ are independent standard Brownian motions. In Fig.~\ref{fig:1}, we use $\tau = 10 \,\text{ms}$.

\end{document}